\documentclass[aps,twocolumn,prl,showpacs,superscriptaddress,floatfix]{revtex4}

\usepackage{amssymb,amsmath}

\begin{document}
%\title{Stability of the Incoherent Phase in a System of Globally Coupled Rotors}
\title{Stability and Ensemble Inequivalence in a Globally Coupled System}

\author{M. Y. Choi}
\affiliation{Department of Physics, Seoul National University, Seoul 151-747, Korea} 
\affiliation{Korea Institute for Advanced Study, Seoul 130-722, Korea}

\author{J. Choi}
\affiliation{Department of Physics, Keimyung University, Taegu 704-701, Korea}

\begin{abstract}
We consider a system of globally coupled rotors, described by a set of
Langevin equations, and examine stability of the incoherent phase. 
The corresponding Fokker-Planck equation, providing a unified description of 
microcanonical and canonical ensembles, bears a few solutions, 
depending upon the ensemble.  It is found that the stability
of each solution varies differently with the temperature, 
revealing the inequivalence between the two ensembles. 
This also suggests a physical explanation of the quasi-stationarity observed in 
recent numerical results.
\end{abstract}


\maketitle

A system of oscillators, coupled sinusoidally with each other, provides 
a prototype model for a variety of oscillatory phenomena in nature. 
With short-range coupling, i.e., nearest-neighbor interactions, 
such an oscillator system describes an array of Josephson junctions which has been 
a subject of extensive studies~\cite{myc}.  
The opposite limit, corresponding to infinite-range interactions,
leads to a system of globally coupled rotors, 
which simulates those systems having long-range forces in physics and biology. 
In spite of the mean-field character of such a system of globally coupled 
rotors, the system turned out to exhibit rich features in dynamical and
statistical properties~\cite{ruf,ano,mta,yos,has,zan}. 
Specifically, one can find an analytic solution in the canonical
ensemble and the system has an equilibrium phase transition at a finite
critical temperature in the ferromagnetic case.  On the other hand,
direct dynamical simulations reveal that it takes quite a long time for 
the system to reach the thermodynamic equilibrium~\cite{ano,yos}. 
This anomalous diffusion phenomenon, dubbed quasi-stationarity, 
is observed only in microcanonical calculation, thus suggesting
that there may exist inequivalence between microcanonical and canonical ensembles. 
Such a quasi-stationary state is believed to survive down well below the
equilibrium critical temperature, attracting recently much
attention together with controversy~\cite{mta,zan}. 

This work is to examine the stability of the incoherent (paramagnetic) phase 
in the system of globally coupled rotors and to provide an explanation
as to the origin of the inequivalence between microcanonical and canonical ensembles.
We consider the Fokker-Planck equation (FPE), which governs the time evolution of
the system in both ensembles and thus provides a unified description, 
and obtain its stationary solutions, depending on the ensemble. 
The stability of each solution is probed and found to behave differently
with the temperature. In particular, the microcanonical ensemble gives rise to additional 
incoherent solutions, which turn out to be neutrally stable even below the 
critical temperature. 
Such neutral stability may give rise to quasi-stationarity, 
thus suggesting a resolution of the controversy. 

The dynamics of the coupled rotor system is governed by
the set of equations of motion for the phase $\phi_i$ of the $i$th rotor:
\begin{equation}\label{eqm}
    M\ddot{\phi_{i}} +\sum_j J_{ij} \sin (\phi_i -\phi_j)=0,
\end{equation}
where $M$ is the inertia of each rotor and $J_{ij}$ represents the coupling
strength between rotors $i$ and $j$.
With the introduction of the canonical momentum $p_i = M\dot{\phi_i}$, 
the above equations are transformed into a set of canonical equations
with the Hamiltonian 
\begin{equation}\label{hamil}
    \mathcal{H}=\sum_{i}\frac{\displaystyle{p^{2}_{i}}}{\displaystyle{2M}}
    -\sum_{i<j} J_{ij}\cos (\phi_i-\phi_j),
  %=K+V ,
\end{equation}
on which the microcanonical description is based.

On the other hand, in the canonical description the system is in contact with
a heat reservoir of temperature $T$ and described, in a most general way, by
the set of Langevin equations:
\begin{equation}\label{eqmo}
    M\ddot{\phi_{i}}+\Gamma \dot{\phi_{i}} +\sum_j J_{ij} \sin (\phi_i
    -\phi_j)=\eta_i ,
\end{equation}
where $\Gamma$ is the damping coefficient 
and the Gaussian white noise $\eta_i (t)$ is characterized by the average
$\langle \eta_i (t) \rangle =0$ and the correlation
$\langle \eta_i (t) \eta_j (t')\rangle = 2 \Gamma T \delta_{ij} \delta (t{-}t')$. 
%Here the bracket denotes the average over stochastic variables. 
%When the mass of the particle is not zero, one can write down the
%equation of motion as the Hamiltonian dynamics,
%\begin{eqnarray}
% \dot{\phi_i}&=&\frac{\displaystyle{p_i}}{\displaystyle{M}}
%\nonumber\\  \dot{p_i}&=&
%-\frac{\displaystyle{\Gamma}}{\displaystyle{M}} p_i -\sum_j J_{ij}
%\sin (\phi_i
%    -\phi_j)+ \eta_i.\label{hamdyn}
%\end{eqnarray}
It is straightforward from Eq. (\ref{eqmo}) to derive
the FPE for the probability distribution $P({\phi_i},{p_i},t)$: 
\begin{eqnarray}
 \frac{\displaystyle{\partial P}}{\displaystyle{\partial t}}
  =&-& \sum_i \frac{\displaystyle{p_i}}{\displaystyle{M}}\frac{\displaystyle{\partial P}}
 {\displaystyle{\partial \phi_i}} 
  + \sum_i\frac{\displaystyle{\partial }}{\displaystyle{\partial p_i}}
 \Big[ \frac{\displaystyle{\Gamma}}{\displaystyle{M}} p_i %\right. 
 \nonumber \\
   & & ~~%\left. 
  - \sum_j J_{ij} \sin (\phi_i-\phi_j)
   + \Gamma T \frac{\displaystyle{\partial }}{\displaystyle{\partial p_i}}\Big]
  P .
\label{fp}
\end{eqnarray}
Note that Eq. (\ref{eqmo}) with $\Gamma$ set equal to zero reduces to Eq. (\ref{eqm}), 
showing that Eq. (\ref{fp}) provides the starting point for both descriptions:
the microcanonical one ($\Gamma =0$) and the canonical one ($\Gamma \neq 0$).
In particular, the stationary solution of Eq. (\ref{fp}) is given by the canonical distribution 
$P^{(0)}({\phi_i},{p_i}) \propto e ^{-\mathcal{H}/T}$ with the very Hamiltonian in Eq. (\ref{hamil})
{\em regardless of} $\Gamma$ being zero or not.
%The overall phase $\theta$ manifesting the U(1) symmetry may be absorbed into the
%definition of $\phi_i$. 

In the case of global coupling $J_{ij} = J/N$, the set in Eq. (\ref{eqmo}) decouples into a
single-particle equation
\begin{equation}\label{seqmo}
    M\ddot{\phi} + \Gamma \dot{\phi}+ J \Delta \sin (\phi-\theta ) =\eta ,
\end{equation}
where the order parameter $\Delta e^{i \theta} \equiv N^{-1}\sum_i^N e^{i\phi_i}$ 
measures the emergence of coherence in the system and
%has been defined by the average over all rotors and
the rotor index $i$ has been suppressed. 
This in turn leads to the FPE for the single-rotor probability distribution $P(\phi, p, t)$:
\begin{equation}\label{fps}
  \frac{\displaystyle{\partial P}}{\displaystyle{\partial t}}
  =- \frac{\displaystyle{p}}{\displaystyle{M}}\frac{\displaystyle{\partial P}}
  {\displaystyle{\partial \phi}}+ \frac{\displaystyle{\partial }}{\displaystyle{\partial p}}
  \left[ \frac{\displaystyle{\Gamma}}{\displaystyle{M}} p +J \Delta
   \sin (\phi -\theta ) + \Gamma T \frac{\displaystyle{\partial }}{\displaystyle{\partial p}}\right] P,
\end{equation}
which reduces, in the absence of damping ($\Gamma =0$), to the FPE for the microcanonical ensemble:
\begin{equation}\label{fpsno}
    \frac{\displaystyle{\partial P}}{\displaystyle{\partial t}}
  =- \frac{\displaystyle{p}}{\displaystyle{M}}\frac{\displaystyle{\partial P}}
  {\displaystyle{\partial \phi}}+
   J \Delta \sin (\phi -\theta ) \,\frac{\displaystyle{\partial P}}{\displaystyle{\partial
   p}}.
\end{equation}
In terms of this probability distribution, the order parameter is simply given by
$\Delta e^{i \theta} =  
%\equiv N^{-1}\sum_i^N \langle e^{i\phi_i}\rangle$ 
\langle e^{i\phi}\rangle$, 
%$\Delta$ defined by
%\begin{equation}\label{op}
% \frac{1}{N}\sum_i^N \langle e^{i\phi_i}\rangle \equiv \Delta e^{i \theta} ,
%\end{equation}
where $\langle \cdots\rangle$ denotes the average over the distribution 
$P(\phi, p, t)$. 
As pointed out for the general case, both Eqs. (\ref{fps}) and (\ref{fpsno}) support the same 
stationary solution $P^{(0)}(\phi,p) = \mathcal{Z}^{-1} e^{-\mathcal{H}/T}$ 
with the single-particle Hamiltonian
\begin{equation}\label{spham}
  \mathcal{H}=\frac{\displaystyle{p^{2}}}{\displaystyle{2M}}- J\Delta\cos \phi ,
\end{equation}
where the overall phase $\theta$ manifesting the U(1) symmetry has been absorbed into the
definition of $\phi$. 
Note that $T$ is given in Eq. (\ref{fps}) but remains arbitrary for Eq. (\ref{fpsno}): 
In the microcanonical ensemble the temperature should be defined as a measure of the average 
kinetic energy according to $\langle p^2 \rangle \equiv MT$.
The partition function is determined by normalization:
$\mathcal{Z} = (2\pi)^{-1}\int dp \int d\phi ~e^{-\mathcal{H}/T} 
              = \sqrt{2\pi MT}I_{0}(x),
$
%\begin{eqnarray}%\label{parf}
%   \mathcal{Z}&=&\int dp~ e^{-p^2 /2MT}\int
%   \frac{\displaystyle{d\phi}}{\displaystyle{2\pi}}~e^{(J\Delta /T) \cos
%   \phi} \nonumber \\&=&(2\pi MT)^{1/2}I_{0}(x)
%\end{eqnarray}
where $I_{0}(x)$ is the  modified Bessel function of the zeroth order with
$x\equiv J\Delta /T$.

This approach based on the FPE thus provides a unified description of 
microcanonical and canonical ensembles, and makes clear that both ensembles generate the 
same equilibrium behavior~\cite{ruf}, described solely by the same stationary distribution 
$P^{(0)}(\phi,p)$. 
Namely, in both ensembles the equilibrium order parameter is given by
\begin{equation*}
    \Delta 
 = \int dp\int \frac{\displaystyle{d\phi}}{\displaystyle{2\pi}} P^{(0)}(\phi,p) e^{i\phi} 
 = \frac{1}{I_0 (x)}\int \frac{\displaystyle{d\phi}}{\displaystyle{2\pi}}\,e^{i\phi} e^{x\cos \phi},
\label{sceq}
\end{equation*}
which, upon expanding $e^{x\cos \phi}$ in terms of the modified Bessel functions, yields
\begin{equation}\label{sceqx}
    \frac{\displaystyle{T}}{\displaystyle{J}}\,x
     =\frac{\displaystyle{I_1 (x)}}{\displaystyle{I_0 (x)}}. 
\end{equation}
This determines whether the system exhibits coherence ($\Delta\neq 0$): 
The ordered phase emerges when $T/J$ is smaller than the
slope of $I_1 (x)/I_0 (x)$ at $x=0$, and the ferromagnetic system ($J>0$) undergoes 
a phase transition at the critical temperature $T_c =J/2$. 
% since for small $x$ (near $T_c$) $I_1 (x) \rightarrow x/2$ and $I_0 (x)\rightarrow 1$. 
%In the case of antiferromagnetic coupling ($J < 0 $), Eq. (\ref{sceqx}) 
%becomes $-Tx/|J| = I_1 (x)/I_0 (x)$, leading to the only solution $x=0$. 
%Accordingly, the antiferromagnetic system is concluded to have no phase transition at
%finite temperatures. 
%Notice that these results are valid in both descriptions. 

We now turn our attention to the stability of the incoherent phase,
which depends crucially on the ensemble employed, as shown below. 
For $\Delta =0$, the stationary solution $P^{(0)}(\phi,p)$ reduces to the 
Maxwell distribution, describing the incoherent phase. 
Unlike Eq. (\ref{fps}), however, Eq. (\ref{fpsno}), the FPE for the microcanonical ensemble,
allows (incoherent) solutions of the general form: $P^{(0)}(\phi,p)=f_0 (p)$, 
an arbitrary function of $p$, including the Maxwell distribution. 
Note that the uniformity in $\phi$ guarantees $\Delta =0$ since $\int d\phi \,e^{i\phi}=0$. 
To probe the stability of this stationary solution, we write
$P(\phi,p,t)=f_0 (p)+f_1 (\phi,p,t)$ and accordingly $\Delta (t)=\Delta_0 +\Delta_1 (t)$ 
[with $\Delta_0 = 0$ and $\Delta_1 (t)=\int dp \int \frac{d\phi}{2\pi}\,f_1 (\phi, p, t) e^{i\phi}$]
in Eq. (\ref{fpsno}), which leads to the stability equation
\begin{equation}\label{stabeq}
    \frac{\displaystyle{\partial f_1}}{\displaystyle{\partial t}}
  =- \frac{\displaystyle{p}}{\displaystyle{M}}\frac{\displaystyle{\partial f_1}}
  {\displaystyle{\partial \phi}}+
   J \Delta_1 \sin \phi \,\frac{\displaystyle{df_0}}{\displaystyle{dp}}. 
\end{equation}
Since $f_1 (\phi,p,t)$ and $\Delta_1 (t)$ are periodic in $\phi$,
one can Fourier decompose them into plane waves:
\begin{eqnarray}\label{fou}
      f_1 (\phi, p, t)&=& \sum_k \int d\omega \,e^{i(k\phi- \omega t)}\tilde{f}_k(p,\omega) \nonumber \\
      \Delta_1 (t)&=& %\int dp \,\int\displaystyle{\frac{d\phi}{2\pi}}\,f_1 (\phi, p, t) e^{i\phi}
       \int d\omega e^{-i\omega t}\int dp \,\tilde{f}_{-1}(p,\omega). 
\end{eqnarray}
Note here that the perturbed order parameter is proportional only to $\tilde{f}_{-1}(p,\omega)$.
Substituting Eq. (\ref{fou}) into Eq. (\ref{stabeq}), one finds the following relations:
\begin{eqnarray}\label{omega}
  0 &=& \left(\omega - \frac{\displaystyle{kp}}{\displaystyle{M}}\right)\tilde{f_k}(p,\omega) \nonumber \\
   & & -\frac{J}{2}\!\left[\int\!dp'\tilde{f}_{-1}(p',\omega)\right]
      \!(\delta_{k,1}- \delta_{k,-1})\,\frac{df_0}{dp} ,
\end{eqnarray}
satisfied by the Fourier coefficients. 
%where $\delta_{i,j}$ is the usual Kronecker symbol. 
Provided $\omega \neq kp/M$, it is obvious that all Fourier coefficients
vanish except those corresponding to $k=\pm 1$. 
When $\omega = kp/M$, we have a continuous spectrum which correspond to a
time-dependent solution of the FPE with $\Delta =0$; this will be discussed later. 
For $k=\pm 1$, we divide Eq. (\ref{omega}) by $(\omega \mp p/M)$ and integrate over $p$, 
to obtain
\begin{equation}\label{f1}
 %   \int dp\tilde{f}_{-1}(p,\omega)+\frac{JM}{2}\int dp \frac{\displaystyle{f'_0 (p)}}
%    {\displaystyle{p+ \omega /M}}\int
%    dp'\tilde{f}_{-1}(p',\omega)=0 \equiv
    [1+\chi (\omega)] \int dp\,\tilde{f}_{-1}(p,\omega)=0,
\end{equation}
where the response function is given by
\begin{equation}\label{chi}
\chi (\omega)=\frac{JM}{2}\int dp \,\frac{\displaystyle{f'_0 (p)}}
    {\displaystyle{p+ \tilde{\omega}}}
\end{equation}
with $f'_0 (p) \equiv df_0 /dp$ and $\tilde{\omega}\equiv M\omega$. 
Accordingly, for a nontrivial solution, we must have
$1+\chi(\omega) =0$. 
%in (\ref{f1}) since $\Delta_1 $ is proportional to $\tilde{f}_{-1}$. 
Since $\chi(\omega)$ in Eq. (\ref{chi}) has a simple pole at $p = -\tilde{\omega}$
on the complex $p$ plane on which the integration is performed,
one needs to extend the frequency to complex values, i.e.,
$\tilde{\omega}=\tilde{\omega}_r +i\tilde{\omega}_i$. 
The appropriate analytic continuation then leads to
the following relation, depending on the imaginary part of $\tilde{\omega}$:
\begin{equation*} 
%\label{chip}
\displaystyle{\frac{2}{JM}} \chi(\omega)
  = \left\{\begin{array}{ll}
     \int_{-\infty}^{\infty} dp \,\frac{\displaystyle{f'_0 (p)}}
    {\displaystyle{p+ \tilde{\omega }}} &~\mbox{for} ~ \omega_i > 0 \\
     \mathcal{P}\int_{-\infty}^{\infty} dp \,\frac{\displaystyle{f'_0 (p)}}
    {\displaystyle{p+ \tilde{\omega }}} -i\pi f'_0 (-\tilde{\omega})&~\mbox{for} ~ \omega_i = 0 \\
     \int_{-\infty}^{\infty} dp \,\frac{\displaystyle{f'_0 (p)}}
    {\displaystyle{p+ \tilde{\omega}}}-2i\pi f'_0 (-\tilde{\omega}) &~\mbox{for} ~ \omega_i < 0 ,
    \end{array} \right.
\end{equation*}
where $\mathcal{P}$ stands for the principal part. 

%Depending on the sign of the imaginary part of $\tilde{\omega}$,
We thus have three different situations to determine the frequency:
When $\omega_i >0$, the perturbation $\Delta_1 (t)$ grows
indefinitely to make the unperturbed solution $\Delta_0 \,(=0)$ unstable. 
In this case the conditions for the real and the imaginary parts of the
frequency to satisfy are
\begin{equation}\label{unsta}
\left\{\begin{array}{ll}
     \displaystyle{\frac{2}{JM}}+\int_{-\infty}^{\infty} dp \,
     \frac{\displaystyle{(p+\tilde{\omega}_r)f'_0 (p)}}
   {\displaystyle{(p+ \tilde{\omega}_r})^2+\tilde{\omega}^2_i}=0 \\
%    \displaystyle{\tilde{\omega}_i}
   \int_{-\infty}^{\infty} dp \,\frac{\displaystyle{f'_0 (p)}}
    {\displaystyle{(p+ \tilde{\omega}_r)^2+\tilde{\omega}^2_i}}=0.
    \end{array} \right.
\end{equation}
In the opposite case ($\omega_i< 0$), the perturbation dies out to make the system stable. 
The condition is then given by
\begin{equation}\label{stab}
\left\{\begin{array}{ll}
    \displaystyle{\frac{2}{JM}}+\int_{-\infty}^{\infty} dp \,
     \frac{\displaystyle{(p+\tilde{\omega}_r)f'_0 (p)}}
    {\displaystyle{(p+ \tilde{\omega}_r)^2+\tilde{\omega}^2_i}}+2\pi \textrm{Im}f'_0 (-\tilde{\omega})=0\\
     \displaystyle{\tilde{\omega}_i}\int_{-\infty}^{\infty} dp \,\frac{\displaystyle{f'_0 (p)}}
    {\displaystyle{(p+ \tilde{\omega}_r)^2+\tilde{\omega}^2_i}}+2\pi
    \textrm{Re}f'_0(-\tilde{\omega})=0.
   \end{array} \right.
\end{equation}
Finally, when $\omega_i =0$, the condition simply reads
\begin{equation}\label{neu}
\left\{\begin{array}{ll}
    \displaystyle{\frac{2}{JM}}+\mathcal{P}\int_{-\infty}^{\infty} dp \,
     \frac{\displaystyle{f'_0 (p)}}{\displaystyle{p+\tilde{\omega}_r}}=0 \\
    f'_0 (-\tilde{\omega}_r)=0,
   \end{array} \right.
\end{equation}
and the system is characterized by the neutral stability, 
oscillating with frequency $\omega_r$. 
When this condition is met, the system may stay in this state rather a long time 
presumably until higher-order nonlinear terms come into play and break
neutrality. 

We are now ready to discuss the stability of the incoherent phase 
associated with various solutions of the FPE: the uniform distribution and the rotating distribution
as well as the Maxwell distribution; the first two exist only for the microcanonical ensemble while
the third for both ensembles.

\emph{Uniform distribution.} In this simplest case, momenta are distributed 
uniformly in the range $[-\alpha,\alpha]$, i.e., $f_0(p)=1/2\alpha$ for $-\alpha<p<\alpha$. 
Known as a water-bag distribution, this has been adopted in most of
dynamical calculations performed so far. 
Substitution of $f'_0 (p)=(2\alpha )^{-1} [\delta(p{+}\alpha)-\delta(p{-}\alpha)]$ into
Eqs. (\ref{unsta}) - (\ref{neu}), depending on the sign of $\omega_i $, 
leads to
\begin{eqnarray}
   \omega_i&=&\pm\sqrt{\displaystyle{{\frac{J}{2M}}-\left({\frac{\alpha}{M}}
   \right)^2}},~~\omega_r =0 ~~\mbox{for} ~ \alpha < \alpha_1 \nonumber \\
    \omega_r&=&\pm\sqrt{\displaystyle{\left({\frac{\alpha}{M}}\right)^2-\frac{J}{2M}}}
        ,~~\omega_i=0 ~~\mbox{for} ~ \alpha > \alpha_1 ,
\end{eqnarray}
which indicates that the frequency of $\Delta_1 (t)$ and $f_1(t)$ has only the imaginary/real part
for $\alpha$ smaller/larger than $\alpha_1 \equiv \sqrt{JM/2}$. 
Recalling the definition of the temperature in the microcanonical description,
we have $T=\langle p^2 \rangle /M=\alpha^2 /3M$, from which we find $T_1 =J/6$. 
At low temperatures ($T < T_1$) the frequency of the perturbation has the imaginary
part, giving rise to the growth (as well as decay) of the perturbation.  
Accordingly the solution is unstable. 
However, at temperatures $T> T_1$, the system oscillates with the (real) frequency,
and neither dissipation nor growth can be observed.
We point out here that $T_1$ above which the neutral stability emerges
is well below the mean field critical temperature $T_c =J/2$. 
%(obtained in both descriptions.)  
This is precisely what has been observed
in various numerical simulations based on the microcanonical ensemble: 
Starting from the incoherent initial configuration ($\Delta_0 =0$) with the 
uniform distribution, many authors reported that the system remains in
the incoherent state for a long time, before it eventually falls into the 
coherent state ($\Delta_0 \neq 0$).
% which was referred as the quasi-stationary state. 
%Although there is some
%controversy on the existence of quasi-stationary state below
%$T_c$,~\cite{zan} many authors report that the system remains in
%the quasi-stationary state for considerably long time before
%eventually fall into the true equilibrium.
Such quasi-stationarity persists down well below $T_c$~\cite{com1},
but the present analysis shows that it ceases to exist below $T_1$. 

\emph{Maxwell distribution.} We next consider the
Maxwell distribution, $f_0 (p)= (2\pi MT)^{-1/2} e^{-{p^2/2MT}}$. 
For $\omega_i > 0$, substituting $f'_0 (p)=-(p/MT)f_0 (p)$ into Eq. (\ref{unsta})
and noting $f'_0 (-p)=-f'_0(p)$, we find that $\omega_r =0$ is clearly a
solution of the second one in Eq. (\ref{unsta}). From the first of Eq. (\ref{unsta}), 
we obtain
\begin{equation}\label{max1}
    1-\frac{2T}{J}=\sqrt{\pi}y \,e^{y^2}\textrm{erfc}(y),
\end{equation}
where $\textrm{erfc}(y)$ is the complementary error function with
$y\equiv \sqrt{M/2T} \omega_i$.  Since the right-hand side is
monotonically increasing from zero to unity as $y$ grows from zero, 
there is one real solution present for $T<J/2$ and no solution for $T>J/2$.
This implies that the system is unstable for $T<J/2$ due to the presence of the solution 
$\omega_i>0$.  
For $\omega_i < 0$, we first note 
%$f'_0 (-\tilde{\omega})$ in Eq. (\ref{stab})
\begin{equation}
       f'_0 (\tilde{\omega}_r {+}i\tilde{\omega}_i)=-\displaystyle{\frac{\tilde{\omega}_r
         +i\tilde{\omega}_i}{\sqrt{2\pi M^3 T^3}}} \,
         e^{-({\tilde{\omega}^2}_r-{\tilde{\omega}^2}_i)/2MT}
            e^{-i{\tilde{\omega}}_r{\tilde{\omega}}_i /MT}, \nonumber
\end{equation}
from which we see $\textrm{Re}f'_0(\tilde{\omega}_r{+}i\tilde{\omega}_i)=0$ 
for $\tilde{\omega}_r=0$ while 
$\textrm{Im}f'_0 (i\tilde{\omega}_i)=-(2\pi M^3 T^3)^{-1/2} \omega_i \,e^{{\tilde{\omega}^2}_i /2MT}$. 
This suggests that $\tilde{\omega}_r=0$ is again a solution of the second equation 
in Eq. (\ref{stab}) whereas the first one reduces to
\begin{equation}
    \frac{2T}{J}-1=\sqrt{\pi}|y| \,e^{y^2}[2-\textrm{erfc}(|y|)]. 
\end{equation}
Similarly to the $\omega_i  > 0$ case, the right-hand side of the above equation
is monotonically increasing with $|y|$, rendering a solution $\omega_i <0$ only 
for $T> J/2$.  Finally, when $\omega_i = 0$, we have also $\omega_r = 0$ from 
$f'_0 (-\tilde{\omega}_r)=0$; this trivial solution appears at $T=J/2$, 
seen from the first one of Eq. (\ref{neu}).
We thus conclude that the Maxwell distribution, leading to the incoherent phase 
at $T > T_c = J/2$, becomes unstable below $T_c$, where coherence develops. 

This conclusion holds also for the canonical ensemble, where the Maxwell distribution 
is the only stationary distribution: Equation (\ref{fps}) leads to the stability
equation in the form of Eq. (\ref{stabeq}) with the additional term depending on $\Gamma$; 
this, however, has no effects on the perturbation in $\phi$, giving rise to 
the same instability below $T_c$. 

\emph{Rotating distribution.} 
%Discussions presented so far was on time-independent solutions of the FPE. 
In addition to the time-independent solutions discussed so far,
Eq. (\ref{fpsno}) also possesses a time-dependent solution 
of the general form $P^{(0)}(\phi,p,t)=u(\phi{-}pt/M)$ for $\Delta =0$. 
In this rotating solution the phase grows continuously with a continuous spectrum as
mentioned previously. 
Requiring periodicity in $\phi$, we write
\begin{equation}
    P^{(0)}(\phi,p,t)=\sum_k e^{ik(\phi -\frac{p}{M})t}F_k (p),
\end{equation}
where $F_k (p)$ is an arbitrary function except for $F_{\pm 1}(p)=0$ 
due to $\Delta =0$.  
The stability analysis for this solution is entirely similar to
that of stationary solutions and will not be repeated here. 
%We find similar expression as in Eq. (\ref{f1}) with $\chi(\omega)$ as
The response function is given by 
\begin{equation}\label{}
    \chi (\omega)=\frac{JM}{2}\int dp \left[\frac{\displaystyle{F'_0 (p)}}
    {\displaystyle{p+ \tilde{\omega}}}-\frac{\displaystyle{F'_{-2} (p)}}
    {\displaystyle{p- \tilde{\omega}}}\right]
\end{equation}
in place of Eq. (\ref{chi}), and appropriate changes should be made in the subsequent 
equations %from (\ref{chip}) 
through (\ref{neu}). 
Although one can consider various distributions for $F_k$~\cite{cc}, 
we only quote the results from uniform distributions for $F_0 (p) $ and $F_{-2}(p)$, 
with the distribution width $2\alpha$ as before:
\begin{eqnarray}
     \omega_i& =&\pm\sqrt{\displaystyle{{\frac{J}{M}}-\left({\frac{\alpha}{M}}
   \right)^2}},~~\omega_r =0
      ~~\mbox{for}~ \alpha < \alpha_2 \nonumber \\
    \omega_r
    &=&\pm
    \sqrt{\displaystyle{\left({\frac{\alpha}{M}}\right)^2 -{\frac{J}{M}}}}
        ,~~\omega_i=0
        ~~\mbox{for}~ \alpha > \alpha_2
\end{eqnarray}
with $\alpha_2 \equiv J/M$.  Again, one can relate the average kinetic
energy with the temperature: $T = \langle p^2\rangle/M = \alpha ^2 /3M$,
% as
%\begin{eqnarray}
%    \frac{T}{2}=\frac{\langle p^2\rangle}{2M}&=&\sum_k \int dp \,d\phi \frac{p^2}{2M} e^{ik(\phi
%    -\frac{p}{M}t)} F_k (p)\nonumber \\
%    &=& \int dp\, \frac{p^2}{2M}F_0 (p) =\frac{\alpha ^2}{6M},
%\end{eqnarray}
from which one has $T_2 =\alpha^2_2 /3M=J/3$. 
Thus the rotating solution has neutral stability for $T> T_2$ and
becomes unstable below $T_2$. Note once again that $T_2$ is lower than the equilibrium
critical temperature $T_c$.
%%%%%
\begin{table}
\begin{tabular*}{8.5cm}{@{\extracolsep{\fill}}|ccccc|}
\hline
& & Maxwell & uniform & rotating\\
\hline
&$J/2< T$& s & n & n \\
& $J/3<T<J/2$& u  & n & n \\
& $J/6<T<J/3$ & u & n & u \\
& $T<J/6$ & u & u & u \\
\hline
\end{tabular*}
\caption{\label{tab} Stability of the incoherent phase for various distributions. 
Whereas the Maxwell distribution corresponds to both ensembles, the other two 
correspond only to the microcanonical ensemble.  
The letters s, n, and u stand for stable, neutrally stable, and unstable, 
respectively.}
\end{table}
%%%%%

Table I summarizes our results of the stability analysis for various solutions, 
depending on the temperature. 
While coherence develops in equilibrium below the transition temperature $T_c =J/2$, 
the dynamic behavior of the incoherent phase depends on the ensemble: 
In the canonical ensemble the Maxwell distribution is the only stationary distribution
corresponding to the incoherent phase and becomes unstable below $T_c$. 
On the other hand, the microcanonical ensemble gives rise to additional distributions 
such as the uniform and the rotating one, which remain neutrally stable 
even below $T_c$. 
Putting these together, we are led to the plausible conclusion that the quasi-stationarity 
found in numerical simulations below $T_c$ has its origin in the neutrally stable
solutions in the microcanonical ensemble and will not appear in the canonical ensemble.

Similar calculations for the antiferromagnetic system (with $J < 0$),
where coherence does not emerge [see Eq. (\ref{sceqx})], 
lead to that all three types of solutions are always neutrally stable \cite{cc}.  
This may give a clue to nature of the non-stationary behavior observed recently \cite{af}.

%In summary, we have studied the
%stability of paramagnetic phases of Hamiltonian mean field model.
%From a single particle equation of motion, with an order parameter
%determined self-consistently, we write down the FPE and find
%stationary as well as time-dependent rotating solutions for
%paramagnetic phase($\Delta_0 =0$). The stability analysis reveals
%that the \emph{stable} stationary solution (Maxwell),
%corresponding to the canonical ensemble, becomes unstable passing
%the equilibrium critical temperature. Meanwhile, solutions valid
%in the microcanonical ensemble only retain the \emph{neutral}
%stability at the same temperature where the same solution became
%already unstable. %until become unstable at low temperatures.
%This neutral stability persists down below
%equilibrium critical temperature, explaining the observed
%quasi-stationarity in numerical experiments.

M.Y.C. thanks the CNRS at Universit\'e Louis Pasteur for hospitality,
where part of this work was accomplished. 
This work was supported in part by the Overhead
Research Fund of SNU and by the Basic Research Program
(Grant No. R01-2002-000-00285-0) of KOSEF.


\begin{thebibliography}{99}
\bibitem{com1}
Numerical experiments usually use the internal energy $U$ as the control parameter,
which is related with the temperature $T$ via $U=T/2 + (1-\Delta^2)/2$ for $J=1$. 
We here use a slightly different definition of the potential energy.
\end{thebibliography}
\end{document}